\documentclass[prd,preprintnumbers,nofootinbib,aps,9pt]{revtex4}
\usepackage{subeqnarray}
\usepackage{epsfig,amsmath,amssymb,mathtools}
\usepackage{fancybox}
\usepackage{enumitem}
\usepackage{mathrsfs}
\usepackage{empheq}
\usepackage{soul,ulem}
\usepackage[usenames,dvipsnames]{color}
\usepackage[pagebackref=true, colorlinks=true]{hyperref}
\definecolor{redish}{rgb}{0.7,0.2,0.0}  
\definecolor{bluish}{rgb}{0.2,0.5,0.8}
\definecolor{MyLightMagenta}{cmyk}{0.3,0.7,0.3,0.3}
\hypersetup{linkcolor=redish,          
                  citecolor=blue,        
                  filecolor=magenta,      
                  urlcolor=bluish}          
\DeclareFontFamily{U}{rsfs}{}         
\DeclareFontShape{U}{rsfs}{m}{n}{<5> rsfs5 <6><7> rsfs7          %
  <8><9><10><10.95><12><14.4><17.28><20.74><24.88> rsfs10}{}     %
\DeclareMathAlphabet{\mathfs}{U}{rsfs}{m}{n}                     %

\newcommand{\ba}{\nopagebreak[3]\begin{eqnarray}}
\newcommand{\ea}{\end{eqnarray}}
\newcommand{\bii}{\begin{itemize}}
\newcommand{\eii}{\end{itemize}}

\def \({\left(}
\def \){\right)}
\def \[{\left[}
\def \]{\right]}

\def\pb#1{\rlap{\lower1.5ex\hbox{$\longleftarrow$}}{#1}}
\def\dpb#1{\rlap{\lower1.5ex\hbox{$\Longleftarrow$}}{#1}}
\def\spb#1{\rlap{\lower1.5ex\hbox{$\leftarrow$}}{#1}}
\def\sdpb#1{\rlap{\lower1.5ex\hbox{$\Leftarrow$}}{#1}}
\begin{document}
\title{Cauchy's logico-linguistic slip, the Heisenberg uncertainty principle and a semantic dilemma concerning ``quantum gravity''}
\author{Abhishek Majhi}%
\email{abhishek.majhi@gmail.com}
\affiliation{Indian Statistical Institute,\\Plot No. 203, Barrackpore,  Trunk Road,\\ Baranagar, Kolkata 700108, West Bengal, India\\}

\begin{abstract}
The importance of language in physics has gained emphasis in recent times, on the one hand through Hilbert's views that concern formalism and intuition applied for outer inquiry, and on the other hand through Brouwer's point of view that concerns intuition applied for inner inquiry or, as I call, self-inquiry. It is to demonstrate the essence of such investigations, especially self-inquiry (inward intuition), I find it compelling to report that a careful analysis of Cauchy's statements for the definition of derivative, as applied in physics, unveils the connection to the Heisenberg uncertainty principle as a condition for the failure of classical mechanics. Such logico-linguistic, or semantically driven, self-inquiry of physics can provide new insights to physicists in the pursuit of truth and reality, for example, in the context of Schroedinger equation. I point out an explicit dilemma that plagues the semantics of physics, as far as general relativity and quantum mechanics are concerned, which needs to be taken into account during any attempt to pen down a theory of ``quantum gravity''.\\\vspace{0.1cm}

\end{abstract}

\maketitle


The significance of the roles of logic and language in physics becomes evident  from the study of Hilbert's sixth problem, namely, {\it Mathematical Treatment of the Axioms of Physics}\cite{hilbertprob,corry2004} and the associated modern research that has germinated from such roots e.g. see ref.\cite{gorban} and the references therein. In modern terminology, as Gorban would write, it is the ``{\it semantics of physics}'' that may hold the key to subtle issues concerning the foundations of physics in the pursuit of truth and reality\cite{gorban}. 
However, when the question of truth arises then it is unavoidable to note, as Brouwer asserted, ``{\it intuititon subtilizes logic}'' and ``{\it denounces logic as the source of truth}''\cite{brouwerconscious}. If I consider Einstein's views\cite{einphreal}, then ``{\it physics constitutes a logical system of thought}'' and ``{\it the justification (truth content) of the system rests in the proof of usefulness of the resulting theorems on the basis of sense experiences, where the relations of the latter to the former can only be comprehended intuitively.}'' Now, the truth content of the system must also depend on the truth content of the parts of the system i.e. the axioms which constitute such a logical system. This, as I understand, raises the concern whether the axioms themselves are truthful expressions of our experiences. In other words, while axiomatizing physics to form a logical foundation of our reasoning, in the pursuit of truth through scientific inquiries, it is also a matter of concern to analyze how truthfully the axioms themselves manifest our experience -- a process which Brouwer might have called ``{\it inner inquiry}''\footnote{The words ``inner inquiry'' was used by Brouwer only once as far as I could find in the literature and that also without any specific demonstration or any further explication. I have referred to Brouwer's works for citation purpose and only to point out that the role of intuition had been investigated earlier. However, as far as my understanding is concerned, this work has nothing to do with the literature related to intuitionistic mathematics.} or, which I call, self-inquiry\cite{ll1,lpp}. For example, Klein's proposal to replace ``{\it a point}'' by ``{\it a small spot}'', as noted by Born on p 81 of ref.\cite{born}, is an act of self-inquiry so as to make the first axiom of geometry, as stated by Euclid\footnote{The first axiom of geometry, as stated by Euclid (see p 153 of ref.\cite{euclid1}), reads as follows: ``{\it A point is that which has no part.}''}, a more truthful expression of our experience which, nevertheless, can only be realized through intuition by being conscious about the fact that the meaning of the words ``a point''(abstract) can not be explained without putting ``a small spot'' (concrete) on the paper\cite{klein}. This is somewhat the opposite of Hilbert's attitude of achieving logical completion by formalizing geometry, without worrying about the truth content of Euclid's first axiom of geometry as far as our experience is concerned\cite{hilbertgeometry}. The modern interpretation of Hilbert's attitude is the denial of ``{\it any need for intuition}'' in the process\cite{gorban}, although Hilbert himself wrote on page no. 2 of ref.\cite{hilbertgeometry}, while categorizing the axioms of geometry into separate groups, that ``{\it Each of these groups expresses, by itself, certain related fundamental facts of our intuition.}'' So, intuition is irremovable, inescapable and indispensable;  Kant would write (as quoted by Hilbert on p 1 of ref.\cite{hilbertgeometry}) that ``{\it All human knowledge begins with intuitions, thence passes to concepts and ends with ideas}''\footnote{In view of Kant's statement I would like to emphasize that the word ``intuition'', in this present discussion, must be taken as a substitute for ``human intellect'' which Brouwer wrote too in ref.\cite{brouwerformint}. The present discussion has a priori nothing to do with the specific literature associated with intuitionistic mathematics and intuitionistic logic.}. However, intuition can be applied in two opposite directions as follows. 
\begin{itemize}
\item One is outward directed, which deals with more and more axioms that are required to formalize our experience in accord with what Hilbert suggested: ``{\it we shall try first by a small number of axioms to include as large a class as possible of physical phenomena, and then by adjoining new axioms to arrive gradually at the more special theories.}''\cite{hilbertprob}
\item The other one is inward directed, which deals with the truthfulness of the expressions of our experience as far as the individual axioms are concerned i.e. the language through which we express our experience comes under scrutiny.
\end{itemize}
  While the path of outward inquiry, that deals with the relations of the axioms, the logical consistency of the axioms and the usefulness of the derived conclusions from an empirical point of view\cite{einphreal, hilbertprob}, is a much trodden pasture in modern science, the path of inner inquiry or self-inquiry has remained hitherto untrodden\footnote{An exception is  ref.\cite{ll1} which deals with self-inquiry that concerns logico-linguistic or semantic issues associate with the foundations of physics.}. Interestingly, alongside the resurgence of an interest in Hilbert's sixth problem\cite{gorban}, there has been a proposal by Gisin to analyze the language of physics from a Brouwerian point of view, albeit devoid of any emphasis on, or demonstration of, inner inquiry\cite{gisin,gisin2}. In a nutshell, the essence of analyzing the language of physics seems to be becoming more important than ever in the context of modern research concerning the foundations of physics.

Here, I shall demonstrate the essence of self-inquiry concerning the foundations of mechanics. To emphasize the worth of such an investigation, let me quote a relevant section from Hilbert's statement of the sixth problem\cite{hilbertprob} as follows.\vspace{0.1cm}

``{\it Important investigations by physicists on \ul{the foundations of mechanics} are at hand;....
	It is therefore very desirable that the discussion of \ul{the foundations of mechanics} be taken up by mathematicians also. Thus Boltzmann's work on the principles of mechanics suggests the problem of developing mathematically \ul{the limiting processes}, there merely indicated, which lead from \ul{the atomistic view} to the laws of motion of continua. Conversely one might try to derive \ul{the laws of the motion of rigid bodies} by a limiting process from a system of axioms depending upon the idea of continuously varying conditions of a material filling all space continuously, these conditions being defined by parameters.}''\vspace{0.1cm}

So, the issues of concern were the foundations of mechanics and specifically to understand the connection between the laws of motion and the atomistic view via a limiting process. Now, as far as the laws of motion of rigid bodies are concerned, we can only write those if and only if we can write down the expressions for the definition of instantaneous velocity, instantaneous acceleration and instantaneous rate of change of momentum. On the other hand, from the modern perspective, ``the atomistic view'' is expressed through quantum mechanics and more specifically characterized by Heisenberg uncertainty principle\cite{heisenberg} and Schroedinger equation\cite{schroedinger} to begin with. Needless to say, it is the definition of derivative, from where basic calculus starts, can be considered as an axiom itself in the process of stating the laws of motion and the differential equation that we know as ``Schroedinger equation''.      

With an aim to add some stimulus to the research investigations concerning the language of physics\cite{corry2004, gorban, gisin, gisin2}, and in order to demonstrate a case of self-inquiry concerning the kind of concerns raised in Hilbert's sixth problem,  I intend to analyze Cauchy's definition of derivative as he put it in his own words in his book\cite{cauchycal}. This is because, on a careful study of how Cauchy defined the notion of derivative in his book\cite{cauchycal}, I find that it is radically different from the standard practice in physics\cite{rudin,apostol,spivak}. Quite remarkably, as I find, a careful exposition of Cauchy's statements unveils a straightforward connection to the Heisenberg uncertainty principle\cite{heisenberg} as a condition for the violation of the laws of motion and therefore, for the invalidity of classical mechanics. While there is a much broader virtue associated with the present discussion from the philosopher's and the logician's perspective, which I have elaborated in ref.\cite{lpp}, I believe that the central issue is worth the attention of the mainstream physics community irrespective of the philosopher's and the logician's assessment of such an inquiry.

Before proceeding, for the ease of understanding of the reader, I may specify the meanings of the following symbols to be used: ``$:=$'', ``$\ni$'' and ``$\Leftrightarrow$'' stand for ``defined as'', ``such that'' and ``equivalent to'', respectively.

{\color{black} Cauchy's exposition of ``derivative of a function'' consists of the following steps of reasoning. On page no. 8 of ref.\cite{cauchycal}, he considered the propositions ``$y=f(x)$'' and ``$y+\Delta y=f(x+\Delta x)$''. 
Here, ``$y$'' and ``$x$'' are, {\it a priori, independent}\footnote{Although Cauchy did not clarify the a priori independence of $y$ and $x$, but it is quite obvious because, in physics, ``$y$'' and ``$x$'' denote quantities with different physical dimensions like length, time, etc. which are denoted by distinct symbols $L, T$, etc. More trivially, I believe, we use distinct symbols ``$y$'' and ``$x$'' when we want to express two a priori independent thoughts or two thoughts which we believe to be a priori independent.} variables which get related by the first proposition. We call  ``$y$'' as a ``function of $x$''. ``$\Delta y$'' and ``$\Delta x$'' are the respective increments in ``$y$'' and ``$x$''. Using these two propositions Cauchy obtained ``$\Delta y=f(x+\Delta x)-f(x)$'' and then he} defined ``derivative of a function'', on page no. 11 of ref.\cite{cauchycal}, as follows:\vspace{0.1cm}

  ``{\it $\cdots$ function $y = f (x)\cdots$ variable $x\cdots$ an infinitely small increment attributed to the variable produces an infinitely small increment of the function itself. $\cdots$ set $\Delta x = i$, the two terms of the ratio of differences
\begin{eqnarray}
	\frac{\Delta y}{\Delta x}=\frac{f(x+i)-f(x)}{i}.\label{cder}
\end{eqnarray}	
	will be infinitely small quantities. $\cdots$ these \ul{two terms} indefinitely and simultaneously will approach the limit of zero, the ratio itself may be able to converge
	toward another limit,...}''\vspace{0.1cm}

Cauchy's ``infinitely small quantities'', in modern mathematical notation, appears as ``$\Delta x\to 0, \Delta y\to 0$'' in the above scenario. Therefore,  according to Cauchy's prescription, {\color{black}{\it both ``$\Delta y\to 0$'' and ``$\Delta x\to 0$'' need to hold so that the derivative is definable}}. Thus, from expression (\ref{cder}), considering a truthful conversion of verbal statements into mathematical notations, {\bf I assert} the following:
\begin{align}
\boxed{\text{\bf Cauchy's definition:~~~}\frac{dy}{dx}:=\lim\limits_{\substack{\Delta y\to 0\\ \Delta x\to 0}}\frac{\Delta y}{\Delta x}\quad\ni~ y=f(x),~ \Delta y=f(x+\Delta x)-f(x).\label{cderp}}
\end{align}
 I have skipped the unnecessary step of setting $\Delta x=i$. In the {\bf standard practice}\cite{rudin,apostol}, one finds the following:
\begin{align}
	\boxed{\text{\bf Cauchy's definition:~~~}\frac{dy}{dx}:=\lim\limits_{\substack{\Delta x\to 0}}\frac{\Delta y}{\Delta x}\quad\ni~ y=f(x),~ \Delta y=f(x+\Delta x)-f(x).\label{stan}}
\end{align}
Obviously one can note what the difference is in the two scenarios.
\begin{align}
	\boxed{\text{\bf The difference is the omission of ``$\Delta y\to 0$''.}\label{diff}}
\end{align}
{\color{black} At this point one can certainly object that there is no necessity of separately declaring ``$\Delta y\to 0$'' because it is obvious from the definition of $\Delta y~$!\footnote{The author thanks B. Juarez Aubry for pointing out this issue.} That is, for ``$\Delta x\to 0$'', we have ``$f(x+\Delta x)\to f(x)$'' and then obviously ``$\Delta y\to 0$'' because ``$\Delta y:=f(x+\Delta x)-f(x)$''. However, then such an objection should stand against Cauchy as well for explicitly writing ``{\it two terms}....will approach the limit of zero''. In any case, if we are to decide whether ``$\Delta y\to 0$'' should be explicitly written or not, then it is best to analyze whether there is any particular use or significance of such writing. My motto is to precisely analyze the consequence of overlooking the fact ``$\Delta y\to 0$'' in basic physics so as to convey the importance of declaring facts in symbolic terms is as essential as doing it verbally i.e. the correspondence between {\it object-language} (symbolic form) and the {\it metalanguage} (verbal form, English in the present case)  must be as precise and rigorous as possible (e.g see page no. 20 of ref.\cite{fraenkel}).}

Since I have not found any instance in the literature where Cauchy explicated the importance of such an issue, I consider (\ref{diff}) as {\bf Cauchy's logico-linguistic slip}, {\color{black} which should be considered as a nomenclature of a fact rather than a negative critic of the respected gentleman}.  I have used the adjective ``logico-linguistic'' because I am scrutinizing {\it how reasonably  the mathematical expressions convey the sense carried by the verbal expressions} i.e. {\it reasoning} and {\it language} are the matters of concern\footnote{One can alternatively use the adjective ``semantic'' instead of ``logico-linguistic'' in the present context. However, I prefer ``logico-linguistic'' over ``semantic'' because the former coveys the meaning more explicitly. That is, the former is semantically simpler than the later.}. 

 In any case, the essence of this adjective will appear to be more justified as I proceed to explain the  significance of ``$\Delta y\to 0$'' which can be elucidated through the following conversation {\color{black} where I shall consider $y$ and $x$ (hence, $\Delta y$ and $\Delta x$) as independent entities by ignoring any possible proposed interrelation}.\vspace{0.1cm} 

{\bf Question:} {\it Why does it matter if  ``$\Delta y\to 0$''  is not written in the mathematical expression and how does it affect our understanding of physics?} \vspace{0.1cm}


{\bf Answer:} {\it Writing ``$\Delta y\to 0$'' matters because, if it is not written then the following three problem arises.
	\begin{enumerate}
\item The ratio $\Delta y/\Delta x$ diverges as $\Delta x\to0$, rather than converging to some limit.
\item The mathematical expression is not a truthful conversion of the verbal statements of Cauchy.
\item The physicist does not realize why in some situations classical mechanics fails. 
	\end{enumerate}}
While the first two of the above three assertions can be immediately verified, {\it the third assertion needs more elaboration which I provide as follows}. The physicist deals with {\it quantities} which are expressed in terms of some standard quantity, called {\it unit}, of the same {\it physical dimension} e.g. length is expressed in terms of length unit like meter, kilometer, etc.; time is expressed in terms of time unit like second, microsecond, etc.\cite{bipm}. Let me denote the chosen units of length and time as $\lambda_0$ and $T_0$ respectively i.e. $\lambda_0$ stands for meter, kilometer, etc. and $T_0$ stands for second, microsecond, etc. I write the displacement (of an object as a whole) as $\Delta x= \Delta n_x\lambda_0$ and the time lapse during this displacement as $\Delta t=\Delta n_t T_0$. Further, I call such expressions as ``physico-mathematical'' due to the involvement of physical dimensions alongside the numbers $\Delta n_x, \Delta n_t$. Now, I note the following inter-conversion between verbal statements and physico-mathematical expressions as follows:
\begin{eqnarray}
	\text{\footnotesize ``$\Delta x$ is an infinitesimally small length compared to the length unit $\lambda_0$''}&\Leftrightarrow&\Delta x\lll\lambda_0~~\Leftrightarrow~~\Delta n_x\lll 1,~~\label{delx}\\
	\text{\footnotesize ``$\Delta t$ is an infinitesimally small time compared to the time unit $T_0$''}&\Leftrightarrow&\Delta t\lll T_0~~\Leftrightarrow~~\Delta n_t\lll 1.~~\label{delt}
\end{eqnarray}  
 I must write ``$\Delta n_x\to 0, \Delta n_t\to 0$'' instead of  ``$\Delta n_x\lll 1, \Delta n_t\lll 1$'' in accord with the currently accepted standard notation (however, see ref.\cite{ll1} for an explanation of how the use of ``$0$'' can lead to logico-linguistic fallacy). So, I adopt the following convention of writing:
\begin{eqnarray}
&&\Delta n_x\lll 1~~\Leftrightarrow~~\Delta n_x\to 0,\label{delx1}\\
&&\Delta n_x\lll 1~~\Leftrightarrow~~\Delta n_t\to 0\label{delt1}.
\end{eqnarray}  
Therefore, under the validity of the conditions (\ref{delx}) and (\ref{delt}), and using the notation adopted in (\ref{delx1}) and (\ref{delt1}), instantaneous velocity can be defined, according to Cauchy's definition of derivative as follows:
 \begin{eqnarray}
\boxed{\frac{dx}{dt}:=\lim\limits_{\substack{\Delta n_x\to 0\\ \Delta n_t\to 0}}\frac{\Delta x}{\Delta t}=\(\lim\limits_{\substack{\Delta n_x\to 0\\ \Delta n_t\to 0}}\frac{\Delta n_x}{\Delta n_t}\)\frac{\lambda_0}{T_0}=n_v v_0 ~\ni  n_v:=\lim\limits_{\substack{\Delta n_x\to 0\\ \Delta n_t\to 0}}\frac{\Delta n_x}{\Delta n_t}~\& ~v_0:=\frac{\lambda_0}{T_0}.}
\end{eqnarray}
Now, to write down the laws of motion, the derivative of momentum needs to be defined according to Cauchy's prescription. So, let me consider momentum $(p)$ to be a quantity in its own right (which is sufficient for the present purpose). I denote a change in $p$ as $\Delta p=\Delta n_p p_0$, where $p_0$ is the momentum unit in terms of which $\Delta p$ is expressed. Then the instantaneous rate of change of momentum can only be defined when the following condition is also fulfilled alongside condition (\ref{delt}):
\begin{eqnarray}
	\text{\footnotesize ``$\Delta p$ is an infinitesimally small momentum compared to the momentum unit $p_0$''}&\Leftrightarrow&\Delta p\lll p_0~~\Leftrightarrow~~\Delta n_p\lll 1.~~\label{delp}
	\end{eqnarray}
Then, adopting the usual convention of notation, I may write $\Delta n_p\to 0~~\Leftrightarrow~~\Delta p\lll p_0$, 
from which the definition of instantaneous rate of change of momentum can be written from Cauchy's prescription:
\begin{eqnarray}
	\boxed{ \frac{dp}{dt}:=\lim\limits_{\substack{\Delta n_p\to 0\\ \Delta n_t\to 0}}\frac{\Delta p}{\Delta t}=\(\lim\limits_{\substack{\Delta n_p\to 0\\ \Delta n_t\to 0}}\frac{\Delta n_p}{\Delta n_t}\)\frac{p_0}{T_0}=n_F F_0 ~\ni  n_F:=\lim\limits_{\substack{\Delta n_p\to 0\\ \Delta n_t\to 0}}\frac{\Delta n_p}{\Delta n_t}~\& ~F_0:=\frac{p_0}{T_0}.}
\end{eqnarray}
Now, I may assert that the verbal statements of the laws of motion can only be expressed in physico-mathematical terms to do further calculations if and only if instantaneous velocity and instantaneous rate of change of momentum can be defined according to Cauchy's prescription. In view of this, I may conclude that the conditions (\ref{delx}), (\ref{delt}) and (\ref{delp}) need to remain valid so that classical mechanics is applicable. Consequently, I can assert that the following derived condition needs to hold for the laws of motion and hence, classical mechanics to be applicable:
\begin{eqnarray}
\Delta x. \Delta p\lll L_0 ~~\ni~L_0:=\lambda_0 p_0,\
\end{eqnarray}
which is obtained from (\ref{delx}) and (\ref{delp}) and $L_0$ is an angular momentum unit.  Then, I may now conclude that the laws of motion, and hence classical mechanics, fail when either or both of (\ref{delx}) and (\ref{delp}) do not hold. This failure can be written as 
\begin{eqnarray}
	\boxed{\Delta x.\Delta p\gtrsim L_0.}\label{up}
\end{eqnarray}
Here, I have assumed that the condition (\ref{delt}) remains valid and did not bring it into discussion so that the focus remains only on $\Delta x$ and $\Delta p$ because these quantities appear as the numerators in the definitions of instantaneous velocity and instantaneous rate of change of momentum, respectively. In view of this, I hope, I have been able to explain the significance of ``$\Delta y\to 0$'' in the context of physics and, in the process, I have justified my third assertion regarding how the physicist fails to realize the limitation of classical mechanics owing to Cauchy's logico-linguistic slip.

From the modern standpoint, the identification of ``$L_0$'' with the Planck constant ``$h$''\cite{planck} can be investigated upon by extending such inquiry further as follows. The time independent and the time dependent Schroedinger equations for a massive free particle in one spatial dimension are written as follows, respectively: 
  \begin{eqnarray}
   -\frac{\lambda_c^2}{8\pi}\frac{\partial^2 \psi}{\partial x^2}&=&\psi~~\ni \lambda_c=\frac{h}{mc}, E=mc^2;\label{tise}\\
 \frac{i\lambda_c^2}{4\pi}\frac{\partial^2 \psi}{\partial x^2}&=&\tau_c\frac{\partial \psi}{\partial t}~~\ni \lambda_c=\frac{h}{mc}, c\tau_c=\lambda_c.\label{tdse}
  \end{eqnarray} 
 $m=n_m m_0$ is the mass of the particle where $m_0$ is the mass unit in terms of which $m$ is expressed. $c=n_cv_0$ is the velocity of light in vacuum\footnote{One may possibly object to the use of $c$ in the context of ``non-relativistic'' quantum mechanics. To refute such a possible objection I may assert that the measurement of $c$ has nothing to do with the theories of relativity or quantum mechanics -- it is an experimental fact. Now, according to EPR's condition of completeness of a theory, every element of physical reality must have a counterpart in a physical theory and any physical quantity measured with certainty without disturbing the system (which is light) has a corresponding element in physical reality\cite{epr}. Therefore, if Schroedinger's equations are believed to be the foundations of a physical theory, it is through EPR's arguments that the involvement $c$  is justified. Further, the expression ``$E=mc^2$'' which I have used in eq.(\ref{tise}) is just constructed through dimensional analysis.} expressed in terms of the velocity unit $v_0=\lambda_0/T_0$. $h=n_hL_0$ is the Planck constant expressed in terms of the angular momentum unit $L_0$. $\psi$ is called wave function, which is a function of the space and time variables, denoted by ``$x$'' and ``$t$'', respectively. Identification of ``$L_0$'' and ``$h$'' means $n_h=1$ i.e. the units must be chosen accordingly. However, such a choice is associated with the following doubts which ensue from the discussion regarding Cauchy's logico-linguistic slip. To define ``$\partial^2 \psi/\partial x^2$'', at first we {\bf need to define} ``$\partial \psi/\partial x$'' and while doing so, the following two options arise:  
\begin{eqnarray}
	\text{{\bf Option 1:} ``$\Delta x$ is infinitesimally small compared to $\lambda_0$''}&\Leftrightarrow&\Delta x\lll\lambda_0,~~\label{delxqm}\\
	\text{{\bf Option 2:} ``$\Delta x$ is infinitesimally small compared to $\lambda_c$''}&\Leftrightarrow&\Delta x\lll \lambda_c.~~\label{deltqm}
\end{eqnarray}  
Given the choice of units for $n_h=1$ to hold and considering the condition (\ref{up}) as the flag bearer of quantum mechanics, it now becomes an intricate play of reasoning because, along with the possibilities (\ref{delxqm}) and (\ref{deltqm}), I need to take care of the relations among $\Delta p, p_0, mc$. Such reasoning needs to be done  from {\it direct experience} and {\it intuition} in the laboratory while dealing with measuring units and measured quantities, so as to both refine and build upon {\it formalism} through inner/self-inquiry and outer inquiry, respectively. An investigation is necessary for each and every step of the theory which now looks like a language that the experimenter speaks in the laboratory while making measurements in the pursuit of truth and reality. After all, as Peres boldly asserted from his ``{\it pragmatic and strictly instrumentalist}'' viewpoint,  on page no. (xi) of ref.\cite{peres}, ``{\it quantum phenomena do not occur in a Hilbert space, they occur in a laboratory.}''

 Science could have developed in a different way without Cauchy's logico-linguistic slip.  Nevertheless, history of science can not be changed, but lessons can be learned from such logico-linguistic, or semantically driven, self-inquiry\cite{ll1} to do science with better understanding and more refined reasoning. 
It is very much true that we have an ``{\it endless road to rigour}''\cite{gorban}. But the road runs both outward and inward. As far as the outward inquiry is concerned, Poincare did assert, on page no. 6 of ref.\cite{poincare}, that ``{\it it is precisely in the proofs of the most elementary theorems that the authors of classic treatises have displayed the least precision and rigour.}'' However, in view of this very short report regarding Cauchy's logico-linguistic slip and its immediate connection to the foundations of mechanics, that concerned Hilbert among others, I may humbly assert, {\it a la} Poincare, that {\it it is precisely in the semantics of the axioms and definitions that the authors of classic treatises have displayed the least precision and rigour.}

To add to this note, I may conclude with a logico-linguistic analysis of a very specific statement of Schroedinger so as to explicate how, semantically, quantum mechanics and general relativity (geometry) are founded on contradictory axioms. Indeed the issue is  rooted to the axiom of ``a point'' which I have mentioned as an example, at the very beginning, to explain what self-inquiry is. The statement of concern, which Schroedinger wrote in the introductory paragraph of ref.\cite{schroedinger}, is the following: \vspace{0.3cm}

``{\it ... material points consist of, or are nothing but, wave-systems.}''\vspace{0.3cm}

To analyze this statement, which is the founding ``conception'' of quantum mechanics according to Schroedinger, let me first refine it by removing the ornamental phrase ``or are nothing but'' and deal with the following statement considered as an axiom $S$, say.\vspace{0.0cm}

$$S: \textit{Material points consist of wave-systems.}$$\vspace{0.0cm}

As far as my knowledge of English language is concerned, the above sentence means that {\it any} material point consists of wave-systems  and that is why we can  generalize for more than one material point which we write as ``material points''. So, I may write that the following sentence is semantically equivalent to the above statement.  \vspace{0.0cm}

$$S: \textit{A material point consists of wave-systems.}$$\vspace{0.0cm}

Now, the words ``a material point'' are synonymous to ``a point-mass'' in modern terminology and the mechanics that deals with ``a point-mass'' is what we call point mechanics or classical mechanics\cite{goldstein}. So, the above statement can be recast in the modern form as follows:\vspace{0.0cm}

$$S: \textit{A point-mass consists of wave-systems.}$$\vspace{0.0cm}

This statement can be broken into the following two statements which I may consider as axioms individually.\vspace{0.05cm}

$$P: \textit{ A point-mass consists of parts.}  \quad Q: \textit{Each part is called a wave-system.}$$\vspace{0.05cm}

Then, I can write: $$S\equiv P\wedge Q.$$
Here, the symbol ``$\wedge$'' means ``logical conjunction'' i.e. $S$ is true if and only if both $P$ and $Q$ are true.

Now, as I have mentioned earlier, according to the first axiom of geometry stated by Euclid is the following: {\it A point is that which has no part.} So, considering ``which has no part'' and ``does not consist of parts'' to be semantically equivalent in this context, I can write the first axiom of geometry as follows:\vspace{0.0cm}

$$\textit{A point does not consist of parts.}$$\vspace{0.0cm}

Since classical mechanics and hence general relativity is founded on the concept of ``point-mass'', then the founding axiom of general relativity is just a slight modification\footnote{This ``slight modification'' is the consideration of the word ``mass'' alongside the word ``point'' to form the word ``point-mass''. Certainly a question can be raised whether such a modification is slight or not because, semantically, this modification leads to a contradiction. However, then it becomes a question regarding what one means by the word ``mass'' which can give rise to endless debates. When such debates arise regarding any theoretical aspect, then the experimental implementation of the debatable ideas lead to the settlement i.e. what becomes of importance is how we implement an idea or become operational with it. Certainly when experiment comes into question, then there is no place for debate that the words ``point-mass'' can be considered to be justified because we do perform experiments whose data are interpreted in terms of equations which have been written down based on the words ``point-mass'' and not just the word ``point''. } of the above statement that we write as follows:\vspace{0.00cm}

$$\neg P: \textit{A point-mass does not consist of parts.}$$\vspace{0.0cm}

Here, the symbol ``$\neg$'' means ``logical negation''\footnote{Of course, here I have considered classical logic.}. Now, it becomes trivial to see that if we want to axiomatically think, in Hilbert's way, of merging quantum mechanics and general relativity (for whatsoever reasons) in a single theory, then such a theory should be founded on a contradiction: $S\wedge \neg P\equiv P\wedge \neg P\wedge Q$. That is, we can write the following:\vspace{0.0cm}

$$\text{General Relativity}\wedge\text{Quantum Mechanics}\equiv P\wedge \neg P\wedge Q\wedge\text{other axioms}.$$  \vspace{0.0cm}

Thus, I may now assert, borrowing words from Gorban\cite{gorban}, that the ``{\it hidden contradiction}'' is now out of its hideout, at least, from the ``{\it semantics of physics}'' in the pursuit of truth and reality through a ``{\it theory of everything}'' that is supposed to be obtained by merging quantum theory and general relativity. It appears, at least to me, that if  ``{\it the logical clash between General Relativity and Quantum Theory}''\cite{gorban} is to be solved, then the solution can be obtained through self-inquiry, like that of Klein's intuition of replacing the abstract concept of  ``a point'' by the concrete and direct experience of ``a small spot''\cite{dot1,dot2}.

While this is the scenario associated with the semantics of ``point-mass'', a further query arises in association with light (mass less) propagation, where the word ``point-mass'' is replaced by ``photon''. Since photon does not have mass, but it is considered as a point and it does carry energy, I find it very tempting to call it ``point-energy'' from the semantic perspective. The general practice is to generalize the axiom $S$ for photon or point-energy, say $R$:

$$R:  \textit{A point-energy (photon) consists of wave-systems.} $$

Although such an axiom is not stated explicitly, however it is only due to such generalization one can make sense of ``{\it wave function of photon}''\cite{scully,phwf}.  In view of this the following logical dilemma arises.

Can we explain energy propagation along {\it a null geodesic} (single ray) without using the formula $E=h\nu$, or semantically speaking, without using the concept of ``photon''? Considering a more specific context, can gravitational {\it red}-shift\footnote{I may clarify here that red-shift and time delay experiments both are founded on the same equation. The interpretations are different. What I have questioned here is the theoretical explanation that we give to energy propagation along a single null geodesic. The word ``energy'' is important here because it is to explain this particular term, we need its relation to frequency. This is related to the {\it quality} of the light signal i.e. we are worried about the colour of the light signal. When we explain the gravitational time delay, we are not worried about the colour of the signal, but the time delay between emission and reception of the signal i.e. we only care about the {\it quantity} that we call ``signal''. However, the light signal is always associated with some colour i.e. we can not completely isolate quality and quantity which a fact of the reality of our experience. It is only to interpret the qualitative aspect that we need the involvement of $h$, and hence quantum theory, within general relativity. Therefore, the logical clash arises when we try to look at the qualitative aspect of light propagation.} be tested without using the formula $E=h\nu$, or semantically speaking, without using the concept of ``photon''? If yes, then which is the experiment? If no, then how can general relativity be regarded as completely ``classical'' so that it can be ``quantized'' or can be merged with ``quantum theory''? e.g. see ref.\cite{carlip}. In a nutshell, if we can not avoid using the formula $E=h\nu$ or the word ``photon'' in general relativity, then the theory contains three fundamental constants: (i) the gravitational constant $(G)$ (ii) the velocity of light in vacuum $(c)$ (iii) the Planck constant $(h)$. Therefore, there is an inherent dilemma within general relativity: {\it Is general relativity (gravity) completely classical or does it involve quantum theory?} I believe that such a dilemma is worth an investigation from Hilbert's axiomatic perspective before one attempts to write some theory of ``quantum gravity'' or some ``theory of everything''\cite{carlip}.\\


	{\bf Declarations}\\
	
	{\it Funding:} This work has been supported by the Department of Science and Technology of the Government of India through the  INSPIRE Faculty Fellowship, Grant no.- IFA18-PH208. 
	



\begin{thebibliography}{3}
\bibitem{hilbertprob} D. Hilbert, {\it Mathematical Problems}, Bull. Amer. Math. Soc. 8(10): 437-479 (1902); \href{https://projecteuclid.org/journals/bulletin-of-the-american-mathematical-society-new-series/volume-8/issue-10/Mathematical-problems/bams/1183417035.full}{https://projecteuclid.org/journals/bulletin-of-the-american-mathematical-society-new-series/volume-8/issue-10/Mathematical-problems/bams/1183417035.full}.

\bibitem{corry2004}L. Corry, {\it David Hilbert and the Axiomatization of Physics (1898 - 1918) - From Grundlagen der Geometrie to Grundlagen der Physik}, Springer (2004).


\bibitem{gorban}A. N. Gorban, {\it Hilbert’s sixth problem: the endless road to rigour}, Phil. Trans. R. Soc. A volume 376, issue 2118, 20170238 (2018), \href{https://doi.org/10.1098/rsta.2017.0238}{https://doi.org/10.1098/rsta.2017.0238}, \href{https://arxiv.org/abs/1803.03599}{https://arxiv.org/abs/1803.03599}.
	
\bibitem{goldstein} H. Goldstein, C. P. Poole, J. L. Safko, {\it Classical Mechanics}, Addison Wesley (2001).

\bibitem{cauchycal} D. M. Cates, {\it  Cauchy's Calcul Infinitesimal}, Springer International Publishing (2019).

\bibitem{bipm} {\it BIPM: The International System of Units (SI)}, Brochure, 9th Edition (2019), \href{https://www.bipm.org/documents/20126/41483022/SI-Brochure-9-EN.pdf/2d2b50bf-f2b4-9661-f402-5f9d66e4b507?version=1.9&download=true}{https://www.bipm.org/documents/20126/41483022/SI-Brochure-9-EN.pdf/2d2b50bf-f2b4-9661-f402-5f9d66e4b507?version=1.9\&download=true}.

 \bibitem{rudin} W. Rudin, {\it Principles of mathematical analysis}, McGraw-Hill (1976).


\bibitem{apostol} T. M. Apostol, {\it Calculus - Vol.1, One-Variable Calculus, with an Introduction to Linear Algebra (2nd ed.)}, John Wiley and Sons (1967).


\bibitem{spivak} M. Spivak, {\it Calculus}, 4th Edition, Publish or Perish (2008).

\bibitem{ll1} A. Majhi, {\it A Logico-Linguistic Inquiry into the Foundations of Physics: Part 1}, Axiomathes (2021), \href{ https://doi.org/10.1007/s10516-021-09593-0}{https://doi.org/10.1007/s10516-021-09593-0}.

\bibitem{lpp} A. Majhi, {\it Logic, Philosophy and Physics: A Critical Commentary on the Dilemma of Categories}, Axiomathes (2021), \href{https://doi.org/10.1007/s10516-021-09588-x}{https://doi.org/10.1007/s10516-021-09588-x}.

\bibitem{fraenkel} A. A. Fraenkel, Y. Bar Hillel, A. Levy, {\it Foundations of Set Theory - Studies in Logic and The Foundations of Mathematics, Volume 67}, Elsevier (1973).


\bibitem{poincare} H. Poincare, {\it Science and Hypothesis}, The Walter Scott Publishing Co. Ltd., New York (1905).



\bibitem{heisenberg} See eq.(1) on page no. 14 of the following book: W. Heisenberg, {\it The Physical Principles of Quantum Theory}, Dover Publications (1930). [English translation by C. Eckart, F.C. Hoyt]

\bibitem{schroedinger} E. Schroedinger, {\it An Undulatory Theory of the Mechanics of Atoms and Molecules}, 
Phys. Rev. 28, 1049  (1926); \href{https://doi.org/10.1103/PhysRev.28.1049}{https://doi.org/10.1103/PhysRev.28.1049}.

\bibitem{planck} M. Planck, {\it The Theory of Heat Radiation}, Philadelphia-P. Blakiston's Son \& Co., 
  (1914). [English translation by M. Masius]


\bibitem{peres} A. Peres, {\it Quantum Theory: Concepts and Methods}, Kluwer Academic Publishers (2002).





\bibitem{brouwerformint}L. E. J. Brouwer, {\it Intuitionism and Formalism}, Inaugural address at the University of Amsterdam, read October 14, 1912.

\bibitem{brouwerconscious} L. E. J. Brouwer, {\it Consciousness, philosophy, and mathematics}, Proceedings of the Tenth International Congress of Philosophy (Amsterdam, August 11–18, 1948), North-Holland Publishing Company, Amsterdam1949, pp. 1235–1249.




\bibitem{gisin} N. Gisin, {\it Mathematical languages shape our understanding of time in physics}, Nat. Phys. 16, 114–116 (2020), \href{http://doi.org/10.1038/s41567-019-0748-5}{http://doi.org/10.1038/s41567-019-0748-5}.

\bibitem{gisin2}N. Gisin, Synthese 199, 13345-13371 (2021); \href{https://philpapers.org/archive/GISIIP-2.pdf}{https://philpapers.org/archive/GISIIP-2.pdf}.

\bibitem{euclid1}Euclid, J. L. Heiberg, R. Fitzpatrick - Euclid's Elements of Geometry (2008).

\bibitem{hilbertgeometry} D. Hilbert, {\it The Foundations of Geometry}, The Open Court Publishing Company - Reprinted Edition (1950); Project Gutenberg eBook:   \href{https://www.gutenberg.org/files/17384/17384-pdf.pdf}{https://www.gutenberg.org/files/17384/17384-pdf.pdf}.

\bibitem{einphreal}A. Einstein, {\it Physics and Reality}, Journal of the Franklin Institute,
Volume 221, Issue 3, Pages 349-382 (1936), \href{https://www.sciencedirect.com/science/article/abs/pii/S0016003236910475}{https://www.sciencedirect.com/science/article/abs/pii/S0016003236910475}.

\bibitem{born} M. Born,  {\it Physics in My Generation}, Springer New York (1968).

\bibitem{klein} F. Klein, {\it Elementary Mathematics from an Advanced Standpoint, Geometry}, Dover Publications (2004).

\bibitem{epr}A. Einstein, B. Podolsky, and N. Rosen, {\it Can Quantum-Mechanical Description of Physical Reality Be Considered Complete?}, Phys. Rev. 47, 777 (1935); \href{https://doi.org/10.1103/PhysRev.47.777}{https://doi.org/10.1103/PhysRev.47.777}.	

\bibitem{dot1} A. Majhi, {\it Contradictions, mathematical science and incompleteness}, \href{https://fqxi.org/community/forum/topic/3475}{https://fqxi.org/community/forum/topic/3475}.

\bibitem{dot2} A. Majhi, {\it Resolving the singularity by looking at the dot and demonstrating the undecidability of the continuum hypothesis}, \href{https://hal.archives-ouvertes.fr/hal-03528767}{https://hal.archives-ouvertes.fr/hal-03528767}.

\bibitem{scully} M. Scully, O. Zubairy, M. Suhail, {\it Quantum optics}, Cambridge University Press (2008).

\bibitem{phwf}I. Biyalinicki-Birula, {\it On the wave function of the photon}, Proceedings of the International Conference ``Quantum Optics III'', Szczyrk, Poland, 1993;  Acta Physica Polonica A, Vol. 86, No. 1-2 (1994).  \href{http://przyrbwn.icm.edu.pl/APP/PDF/86/a086z1p08.pdf}{http://przyrbwn.icm.edu.pl/APP/PDF/86/a086z1p08.pdf}.

\bibitem{carlip}{\it Quantum Gravity - Stanford Encyclopedia of Philosophy}, \href{https://plato.stanford.edu/entries/quantum-gravity/}{https://plato.stanford.edu/entries/quantum-gravity/}.

\end{thebibliography}
\end{document}